# A comprehensive liver CT landmark pair dataset for evaluating deformable image registration algorithms

Zhendong Zhang[1], Edward Robert Criscuolo[1], Yao Hao[2], Deshan Yang[1*]

**Author Affiliations**

1. Department of Radiation Oncology, Duke University, Durham, NC, USA

2. Washington University School of Medicine, St. Louis, MO, USA

**Corresponding Author**

[*]Deshan Yang, PhD,

Professor of Radiation Oncology, Duke University

40 Duke Medicine Circle, 04212, 3640 DUMC, Durham, NC 27710

deshan.yang@duke.edu

# Abstract

**Purpose:**

Evaluating deformable image registration (DIR) algorithms is vital for enhancing algorithm performance and gaining clinical acceptance. However, there's a notable lack of dependable DIR benchmark datasets for assessing DIR performance except for lung images. To address this gap, we aim to introduce our comprehensive liver computed tomography (CT) DIR landmark dataset library. This library is designed for efficient and quantitative evaluation of various DIR methods for liver CTs, paving the way for more accurate and reliable image registration techniques.

**Acquisition and Validation Methods:**

Thirty CT liver image pairs were acquired from several publicly available image archives as well as authors' institutions under institutional review board (IRB) approval. The images were processed with a semi-automatic procedure to generate landmark pairs: 1) for each case, liver vessels were automatically segmented on one image; 2) landmarks were automatically detected at vessel bifurcations; 3) corresponding landmarks in the second image were placed using the deformable image registration method; 4) manual validation was applied to reject outliers and confirm the landmarks' positional accuracy. This workflow resulted in an average of ~68 landmark pairs per image pair, in a total of 2028 landmarks for all 30 cases. The general landmarking accuracy of this procedure was evaluated using digital phantoms. Estimates of the mean and standard deviation of landmark pair target registration errors (TRE) on digital phantoms were 0.64 ± 0.40 mm. 99% of landmark pairs had TREs below 2 mm, and 96% had TREs below 1.5 mm.

**Data Format and Usage Notes:**

All data including image files and landmark information are publicly available at Zenodo (https://doi.org/10.5281/zenodo.10553242). Instructions for using our data and MATLAB code can be found on our GitHub page at https://github.com/deshanyang/Liver-DIR-QA.

**Potential Applications:**

The landmark dataset generated in this work is the first collection of large-scale liver CT DIR landmarks prepared on real patient images. This dataset can provide researchers with a dense set of ground truth benchmarks for the quantitative evaluation of DIR algorithms within the liver.

## 1. Introduction

Deformable image registration (DIR) is an image processing procedure to determine the optimal transformation relationship of two corresponding images.[1] DIR serves as a versatile technique for correlating information from separate image series, enabling extensive advanced radiation therapy (RT) tasks such as target definition,[2,3] dose mapping and accumulation,[4] image segmentation,[5,6] motion estimation,[7-9] and treatment response evaluation.[10] However, the use of DIR for real-world clinical tasks, which requires high geometric precision at the voxel level, is still limited due to the underlying uncertainties and unsatisfying accuracy of current DIR algorithms.[1,11] Understanding and characterizing the uncertainties and errors in DIR is critical for optimizing its accuracy and robustness, which are essential for gaining clinical acceptance.[12,13] However, quantitatively evaluating the accuracy of DIR presents considerable challenges.

DIR accuracy is commonly evaluated by using intensity dissimilarity metrics,[14,15] physical or digital phantoms,[16,17] and anatomical landmarks.[18] Intensity dissimilarity metrics, which estimate the registration performance solely relying on image intensities, are not direct demonstrations of anatomical structure correspondence and, thus are unreliable in representing registration errors. Digital or physical phantoms generated with pre-determined deformations are useful to assess DIR correspondence quantitatively. However, these phantoms commonly fail to characterize the complexity of clinical-level images which exhibit variant image quality, artifacts, and ununiform deformations. Computing target registration error (TRE) using anatomical landmark pairs identified by experts is considered the most objective and reliable way to evaluate DIR accuracy. However, the process of manually identifying landmark pairs is highly labor-intensive and subject to inter-observer variations.[4] In addition, manually labeled landmarks are commonly restricted only to the strongest visible features in the images, e.g., the vessel bifurcations that can be observed on 2D slices, resulting in inadequate quantity and spatial distribution.

While manual landmarking was generally limited to a small subset of images or landmark sets for case-specific studies in DIR algorithm papers, several papers were published focusing on DIR benchmark datasets for thoracic CTs, including the POPI dataset,[19] the DIRLAB COPD gene dataset,[20] and the DIRLAB 4DCT dataset,[21,22] have been published. These datasets featured manually or semi-automatically identified landmark pairs in lung images. The widely recognized DIRLAB 4DCT dataset offers 300 landmark pairs per image pair for public access. These lung CT landmark datasets have been pivotal for advancing DIR algorithm development for lung CTs over the past decade. By far, lung deformable registration has stood out as one of the more feasible deformable registration scenarios, consistently achieving mean TRE below 2 mm.[23] Several studies have also been conducted to enhance the landmarking process by developing tools that can automatically detect a large number of landmarks in lung CT images.[18,24-26] Notably, Fu et al. developed an automatic landmark pairs identification workflow by extracting vascular information from lung CT images.[25] The quantity (up to thousands of landmarks) and accuracy of landmark pairs detected using Fu et al.'s method surpassed the landmarks identified by human experts, achieving improved TREs with a mean and standard deviation of 0.47 ± 0.45 mm compared to the inter-observer uncertainty (~0.89 mm) introduced in DIRLAB 4DCT dataset. This confirmed the potential of automatic detection methods in landmark identification.

While most research efforts have focused on thoracic CT images, there have been much fewer efforts on liver images, despite the clinical significance of liver DIR in supporting surgery planning,[27] radiation therapy,[28] and thermal ablation.[29] Compared to lung images, deformable registration in the liver is much more challenging due to complex motion[30] and a much lower effective contrast-to-noise ratio.[31] Studies of liver DIR algorithms revealed that the current DIR performance was sub-optimal, with an average TRE spanning approximately 4 to 8 mm.[32-35] This applied to even the most widely available commercial DIR solutions.[35] Meanwhile, many of these studies also indicated limitations in terms of the absence of sufficient landmarks for the quantitative evaluation of liver DIR algorithms.[28,32,34,35] Osorio et al.[36] evaluated their liver DIR method using 10 to 15 landmarks automatically detected in each case. Fernandez-de-Manuel et al.[27] examined their multimodal non-rigid registration method by employing only 10 manually annotated landmarks per case across seven CT-MR liver image pairs. Polan et al.[37] used between 4 and 17 landmarks per case to assess their DIR methods for implementing radiation dose-volume response in seven patients. In two recent papers involving the evaluation of liver DIR with a dataset of over 20 image pairs, an average of only five manually selected landmark pairs was employed.[33,35] Additionally, Cazoulat et al[24] reported a method that was able detect relatively more landmarks on contrast-enhanced liver CT images (~32 landmarks pairs per case), but the landmark data was not made publicly available. To the authors' best knowledge, there are currently no larger-scale landmark pair datasets for liver DIR evaluation.

The purpose of this paper is to present the first comprehensive large-scale benchmark dataset library for liver CT images. We developed a semi-automated image processing workflow to identify landmark pairs at blood vessel bifurcations between liver CT image pairs, followed by a rigorous manual validation process. This workflow was applied to build a library consisting of 30 pairs of contrast-enhanced CT images each with variant image quality and containing a substantially large number of vessel-bifurcation landmark pairs. This dataset library offers a comprehensive collection of images and corresponding landmark pairs, exceeding all previous reports by a significant margin,[24,27,33,36] for liver CT DIR evaluation. Our purpose is to promote the development of accurate and robust DIR algorithms for liver CTs.

## 2. Acquisition and validation methods

### 2.1. Data Sources

Forty-nine pairs of contrast-enhanced abdominal-pelvis CT scans were initially obtained from several publicly available image repositories as well as from Barnes Jewish Hospital under institutional review board (IRB) approval. Prior to any analysis and processing, the protected health information of the patient subjects was removed from the headers of the DICOM files.

Among the 49 cases, 32 CT image pairs were obtained from the Anti-PD-1 Immunotherapy Melanoma Dataset (Anti-PD-1_MELANOMA).[38,39] The in-plane resolution ranged between 0.63 x 0.63 mm$^2$ and 0.94 x 0.94 mm$^2$, with a slice thickness of 2.5 mm for 63 images or 1.25 mm for one image. Two image pair cases were selected from the Anti-PD-1 Immunotherapy Lung Dataset (Anti-PD-1_Lung).[40] The spatial resolution was 0.78 X 0.78 X 2.5 mm$^3$ across the four images. One image pair was sourced from The Cancer Genome Atlas Esophageal Carcinoma Collection Dataset (TCGA-ESCA),[41] with a voxel size of 0.84 x 0.84 x 1.25 mm$^3$ for the first CT and 0.75 x 0.75 x 1.25 mm$^3$ for the second CT.

Other 14 cases were acquired at Barnes-Jewish Hospital (BJH) with an IRB approval. Each image pair was from abdominal or pelvic cancer patients who had completed their treatments and had multiple CT scans during their courses of diagnosis and treatments. The in-plane image resolutions ranged between 0.61 and 0.98 mm and the slice thicknesses were between 0.7 and 0.98 mm.

Following the implementation of the landmark detection process, 19 of the 49 image pairs were excluded due to poor liver vessel contrast and lack of detectable landmarks. Eventually, a total of 30 pairs of images were kept, all of which contained dense sets of identifiable landmark pairs.

### 2.2. Landmark Pair Identification

#### 2.2.1. General Workflow

Considering the labor-intensive nature and human-observer uncertainty of the manual landmark selection process, a semi-automatic landmark pairs identification procedure was developed. This procedure was designed to efficiently detect landmarks at vessel bifurcations. Such bifurcations were considered because they are repeatable, with reliable positional accuracy between a pair of two intra-patient liver images, and verifiable by human observers.

The landmark pair identification workflow is illustrated below in Figure 1. For each image pair, 1) the liver was segmented for both images. The images were also preprocessed and denoised. 2) In the image with superior quality, designated as image 1, the liver vessel was automatically segmented. 3) Landmarks, represented by vessel bifurcations, were then automatically detected on the skeleton of the segmented vascular tree. 4) The corresponding landmarks in the other image of the image pair, designated as image 2, were placed directly by applying deformable registration between the two images. 5) To account for any uncertainties introduced during the automation process, manual validation was performed to reject outliers and confirm landmarks' position in both images. The results are a large set of well-positioned and manually verified landmark pairs between each image pair. The details of each step will be explained in the following sections.

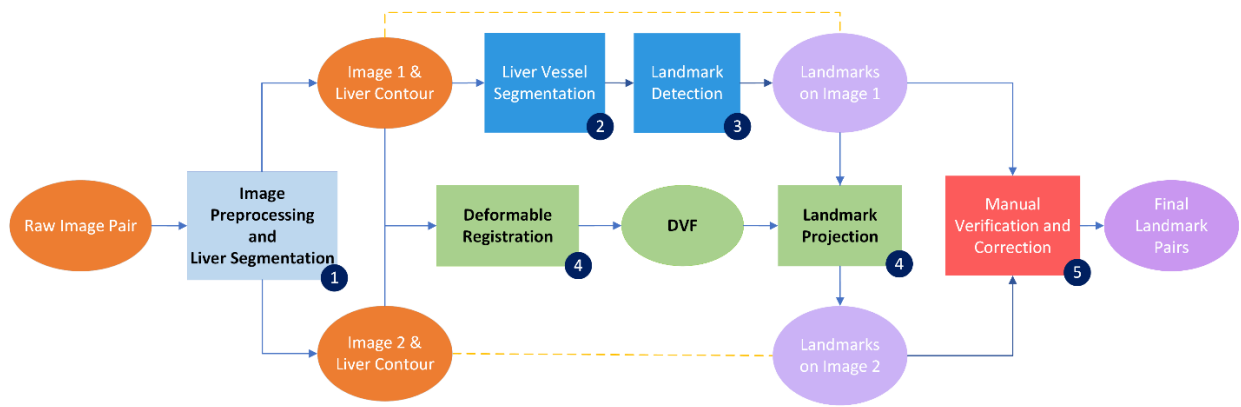

*Figure 1: Landmark pair identification workflow. The workflow contains five major steps which are marked by dark circles. Details of each processing step are described in the following sections.*

### 2.2.2. Image Preprocessing and Liver Segmentation

The abdominal-pelvis images used in this study were originally generated under different scanning conditions, with various image resolutions, image sizes, and image quality. Upon acquiring the images, the liver was automatically segmented in all CT images using DirectORGANS[42] and then manually checked. Before any further processing, all images were up-sampled longitudinally to reduce the slice thickness so that the vascular tree could be generated smoothly in later steps. To facilitate the image registration process, the two images of each case were also cropped to the same image sizes based on the liver contours plus about 10 mm margins. The preprocessed images were then denoised by FFDNet,[43] a fast and flexible denoising neural network with the ability to handle a variant range of noise levels. The denoising strength for each case was manually adjusted based on the level of image quality. Given the challenges of motion blurry and inconsistent voxel intensities of vessels to background tissues associated with liver images, the process of denoising played a crucial role in enhancing the quality of vessel segmentation. This was achieved by effectively reducing unwanted noise and artifacts present in the images. During the process of adjusting the denoising strength, image quality was manually assessed. For each case, the image with higher quality was identified as superior and designated as image 1 in Figure 1. The landmarks detected in image 1 were subsequently designated as the initial landmarks, a process further detailed in section 2.2.3. Examples of denoised liver images are shown in Figure 2.

### 2.2.3. Liver Vessel Segmentation and Landmark Detection

To segment the liver vessel within each image, we first generated a vesselness map using the Frangi vesselness filter,[44] which assigned a likelihood of each voxel being part of the blood vessel. This vesselness map was then employed in segmenting the vessel tree by conducting hysteresis thresholding.[45] The hysteresis thresholding performed edge detection with two thresholds by considering both intensity values and the vesselness and thus was more robust to inconsistent image intensities between liver tissues and vessels than single value thresholding. The thresholding values were determined empirically based on the maximum and average intensity values within the live contour.[45] Examples of vessel segmentation results are shown in Figure 2.

Subsequently, the vessel tree was skeletonized[46,47] and the bifurcation points were directly detected as landmarks on the vessel tree skeleton. The quality of all bifurcation points was evaluated and ranked based on the vesselness value at the points. To avoid repetitive detection of landmarks resulting from imperfect skeletonization, all bifurcation points with a suboptimal ranked quality within a 5 mm distance were considered duplicates and discarded. The quantity of initially detected bifurcation points varies from ~60 to ~160 for different images depending on the image quality and the strength of the CT contrast. All these steps, including vessel segmentation and landmark detection were automated for efficient and repeatable processing. The example of landmark detection results is shown in Figure 2.

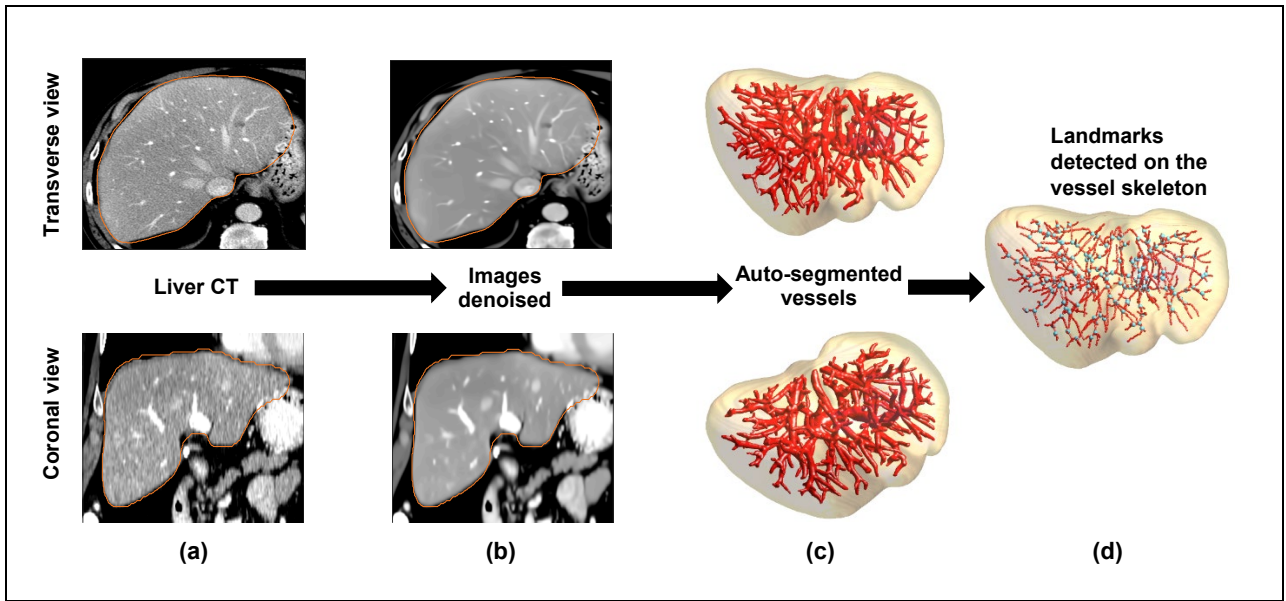

*Figure 2: Examples of image denoising, vessel segmentation and landmark detection results. (a) Transverse/coronal views of the original images with auto-segmented liver contours. (b) Images after denoising. (c) Segmented liver vessels. (d) Landmarks detected on the vessel skeleton. 173 landmarks were initially detected for this case.*

### 2.2.4. Deformable registration and landmark projection

To identify the corresponding landmarks in image 2 after the landmarks were detected in image 1, these two images were deformably registered using pTVreg.[48] pTVreg is a parametric image registration algorithm that can deal with non-smooth displacement and has shown previously in published results to be one of the most accurate DIR algorithms for lung CTs.[48] The landmarks on image 1 were then projected onto image 2 by using the deformation vector field (DVF) computed by pTVreg. This projection process established the initial landmark pairs between each pair of image 1 and image 2. Examples of deformable registration results and landmark projection are shown in Figure 3 and Figure 4.

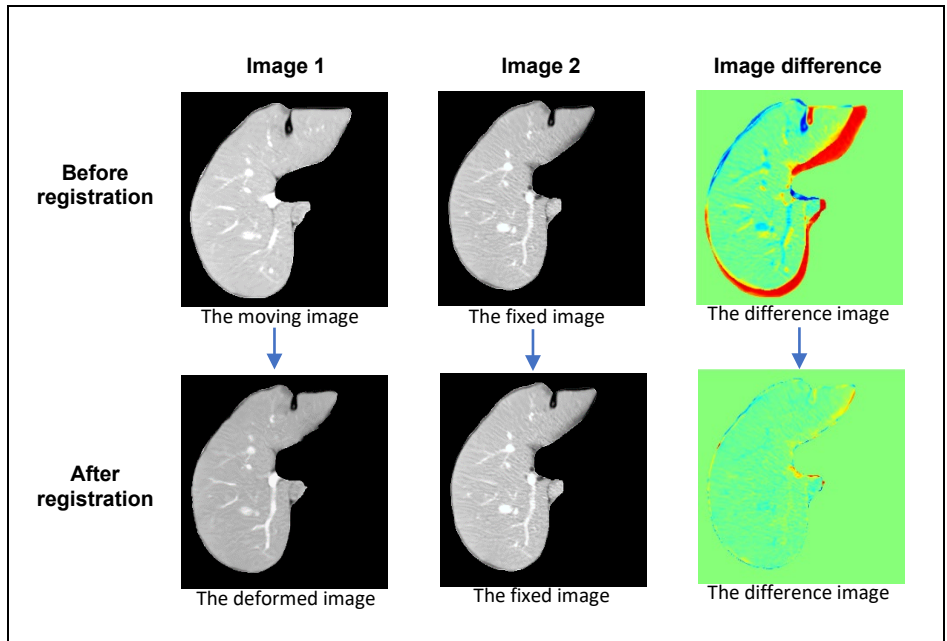

*Figure 3: Examples of deformable registration results. Judging from the appearance of the deformed image and the difference image after registration, image 1 was registered well to image 2.*

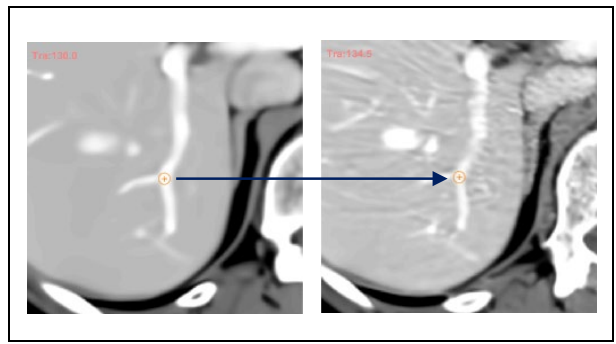

*Figure 4: Examples of landmark projection. The landmark at a vessel bifurcation was detected on image 1 (left) and projected onto image 2 (right) using the computed deformation vector field (DVF).*

### 2.2.5. Landmark Rejection and Manual Verification

The initial landmark pairs might not have been sufficiently accurate due to inconsistent image quality, imperfect liver vessel segmentation, and deformable registration uncertainties. It was imperative to perform a thorough manual verification process to identify and reject falsely detected or projected landmarks and to confirm and possibly manually refine the landmarks' positions. An in-house MATLAB tool was developed to display the vessel segmentation and landmark positions in their original CT images to allow manual confirmation of vessel bifurcation pairing and positions. Rigorous criteria were developed and followed to reject potential outliers. Incorrectly identified landmark pairs primarily fell into four categories: 1) Landmarks detected on very small vessel branches (normally with lengths and diameters within 1 mm). Such branches were highly likely to be incorrectly identified vessels resulting from image noise or inconsistent intensities near the vessel edges. 2) Landmarks visible on image 1 but not visible on image 2. 3) Landmarks detected on incorrectly segmented vessel intersections. 4) Landmarks not projected accurately on image 2.

The manual verification process was executed by two observers independently. Details and examples of manually verified landmarks are exhibited in Figure 5. Overall, ~35% of landmarks were rejected, and all final landmark pairs were confirmed to have a high degree of positional correspondence at clear vessel bifurcations. Throughout the manual verification process, a minimal adjustment of landmark positions was applied to less than ~1% of the landmark pairs. This indicated a high level of positional consistency on the not-rejected landmark pairs. This also suggested that the landmark projection process worked well for the good landmark pairs and had fewer inter-observer variations compared to manual landmark delineation. Eventually, a total of 2028 landmark pairs were identified for 30 image pairs with an average of ~68 pairs for each case.

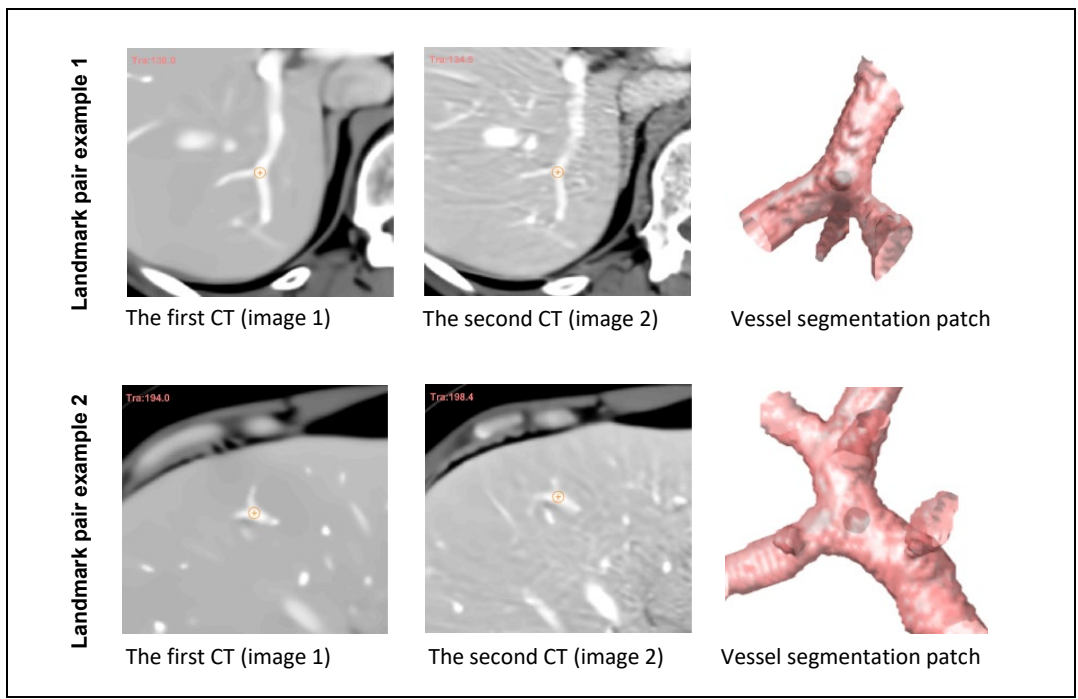

*Figure 5: The landmark pairs are displayed on both CTs as orange color circles and on the 3D local vessel tree as black dots. The manual verification process utilized the displayed images, landmarks, and vessel trees, to confirm good landmark pairs and reject suboptimal ones.*

## 2.3. Assessment of General Landmarking Accuracy

We systematically assessed the general landmarking accuracy of the whole process shown in Figure 1 using digital phantoms. The assessment workflow is shown in Figure 6. An image-to-phantom pair was formed by image 1 of each case and the corresponding phantom image 1' (image 1 deformed by a DVF computed on another patient image pair). This assessment workflow was carried out for all 30 cases, using six different DVFs with one DVF assigned for every five patient cases. Six DVFs were used to avoid bias from one single DVF. Demons, a different DIR algorithm,[49] was used to generate the six DVFs to avoid the bias to the pTVreg algorithm which was used in the landmarking workflow. Landmark pairs were generated between image 1 and image 1' by applying the landmarking workflow shown in Figure 1. The absolute TRE on each landmark pair was computed using the ground truth DVFs. The mean and standard deviation of the TRE were 0.64 ± 0.40 mm for the 30 cases. 99% of landmark pairs had TREs below 2 mm, and 96% of landmark pairs had TREs below 1.5 mm.

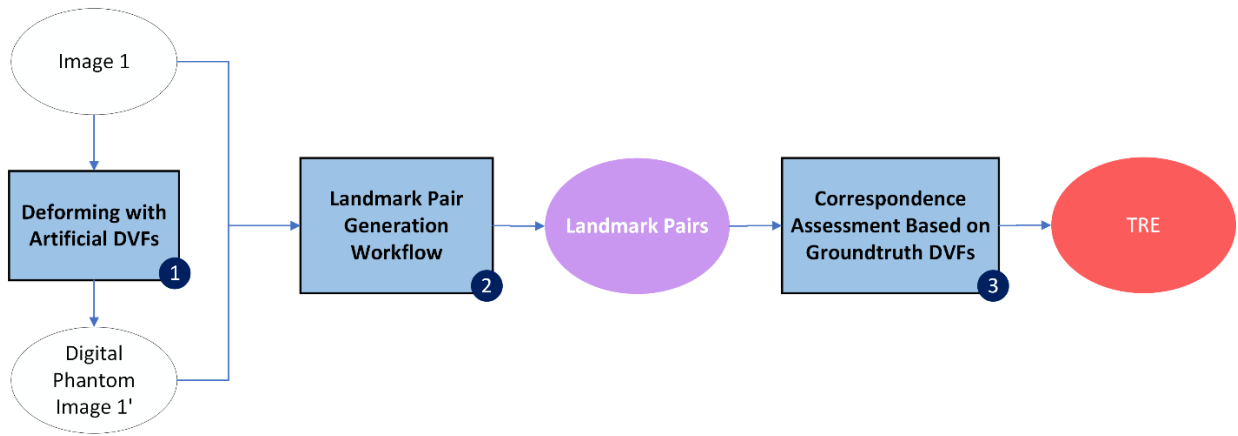

*Figure 6: The landmarking accuracy assessment workflow. In step 1 and step 3, the artificial (ground truth) DVFs were generated by registering other image pairs using a different registration method. Step 2 refers to the workflow in Figure 1.*

## 3. Data format and usage notes

### 3.1. Dataset Overview

The presented liver DIR benchmark dataset comprises 30 pairs of intra-patient abdominal-pelvis CT images, each annotated with a considerable number of well-distributed landmark pairs within the liver. Across the 30 cases, a total of 2028 landmark pairs were labeled, averaging ~68 pairs per case, with a range from 30 to 144. Detailed information about the dataset is provided in Table 1.

The dataset is accessible through Zenodo at https://doi.org/10.5281/zenodo.10553242. A comprehensive usage instruction is available at https://github.com/deshanyang/Liver-DIR-QA. The image data was provided as NIfTI files. The landmark coordinate locations are saved as text files.

*Table 1. List of image profiles and landmark pair numbers. In the data source column, MELANOMA represents Anti-PD-1_MELANOMA dataset; Lung represents Anti-PD-1_Lung dataset; BJH represents Barnes-Jewish Hospital. Scanning parameters are provided as tube kVp, mAs or CT dose index (CTDIvol) values based on information from the image DICOM headers. The mAs*

*values were calculated as tube current · rotation time / pitch.  $^a$ CTDIvol (cGy) values are provided for images missing mAs information in the DICOM headers.*

| Case # | Data source | Image 1 | | Image 2 | | # of landmark pairs |
|---|---|---|---|---|---|---|
| | | Voxel resolution (mm$^3$) | Scanning parameter (kVp / mAs) | Voxel resolution (mm$^3$) | Scanning parameter (kVp / mAs) | |
| 01 | MELANOMA | 0.70 x 0.70 x 2.5 | 120 / 337.9 | 0.70 x 0.70 x 2.5 | 120 / 338.1 | 66 |
| 02 | MELANOMA | 0.82 x 0.82 x 2.5 | 140 / 168.1 | 0.82 x 0.82 x 2.5 | 120 / 28.7 $^a$ | 58 |
| 03 | MELANOMA | 0.70 x 0.70 x 2.5 | 120 / 165.0 | 0.70 x 0.70 x 2.5 | 120 / 191.8 | 91 |
| 04 | MELANOMA | 0.82 x 0.82 x 2.5 | 120 / 285.3 | 0.82 x 0.82 x 2.5 | 120 / 291.8 | 30 |
| 05 | MELANOMA | 0.78 x 0.78 x 2.5 | 120 / 454.3 | 0.82 x 0.82 x 2.5 | 120 / 448.6 | 60 |
| 06 | MELANOMA | 0.70 x 0.70 x 2.5 | 120 / 182.9 | 0.70 x 0.70 x 2.5 | 120 / 174.7 | 38 |
| 07 | MELANOMA | 0.78 x 0.78 x 2.5 | 120 / 14.9$^a$ | 0.78 x 0.78 x 2.5 | 120 / 353.5 | 64 |
| 08 | MELANOMA | 0.70 x 0.70 x 2.5 | 120 / 243.8 | 0.70 x 0.70 x 2.5 | 120 / 252.6 | 39 |
| 09 | MELANOMA | 0.78 x 0.78 x 2.5 | 120 / 277.3 | 0.78 x 0.78 x 2.5 | 120 / 233.2 | 69 |
| 10 | MELANOMA | 0.74 x 0.74 x 2.5 | 120 / 212.9 | 0.74 x 0.74 x 2.5 | 120 / 315.3 | 65 |
| 11 | MELANOMA | 0.78 x 0.78 x 2.5 | 120 / 264.1 | 0.78 x 0.78 x 2.5 | 120 / 273.1 | 92 |
| 12 | MELANOMA | 0.82 x 0.82 x 2.5 | 120 / 603.8 | 0.82 x 0.82 x 2.5 | 120 / 439.7 | 88 |
| 13 | MELANOMA | 0.78 x 0.78 x 2.5 | 120 / 164.2 | 0.78 x 0.78 x 2.5 | 120 / 162.5 | 77 |
| 14 | MELANOMA | 0.86 x 0.86 x 2.5 | 120 / 445.4 | 0.86 x 0.86 x 2.5 | 120 / 379.5 | 68 |
| 15 | MELANOMA | 0.78 x 0.78 x 2.5 | 120 / 429.1 | 0.78 x 0.78 x 2.5 | 120 / 436.4 | 69 |
| 16 | MELANOMA | 0.90 x 0.90 x 2.5 | 120 / 230.0 | 0.90 x 0.90 x 2.5 | 120 / 273.1 | 60 |
| 17 | MELANOMA | 0.70 x 0.70 x 2.5 | 120 / 251.9 | 0.70 x 0.70 x 2.5 | 120 / 268.2 | 83 |
| 18 | Lung | 0.78 x 0.78 x 2.5 | 120 / 203.2 | 0.78 x 0.78 x 2.5 | 120 / 203.2 | 144 |
| 19 | Lung | 0.78 x 0.78 x 2.5 | 120 / 166.6 | 0.78 x 0.78 x 2.5 | 120 / 168.1 | 47 |
| 20 | MELANOMA | 0.86 x 0.86 x 2.5 | 120 / 334.0 | 0.86 x 0.86 x 2.5 | 140 / 309.8 | 50 |
| 21 | MELANOMA | 0.74 x 0.74 x 2.5 | 120 / 213.3 | 0.74 x 0.74 x 2.5 | 120 / 161.7 | 61 |
| 22 | MELANOMA | 0.86 x 0.86 x 2.5 | 120 / 165. 8 | 0.86 x 0.86 x 2.5 | 140 / 168.1 | 114 |
| 23 | MELANOMA | 0.78 x 0.78 x 2.5 | 120 / 164.2 | 0.78 x 0.78 x 2.5 | 120 / 258.4 | 46 |
| 24 | MELANOMA | 0.78 x 0.78 x 2.5 | 120 / 400.7 | 0.78 x 0.78 x 2.5 | 120 / 351.9 | 75 |
| 25 | BJH | 0.89 x 0.89 x 2 | 120 / 18.2$^a$ | 0.61 x 0.61 x 2 | 100 / 20.9 $^a$ | 56 |
| 26 | BJH | 0.87 x 0.87 x 2 | 120 / 20.7$^a$ | 0.98 x 0.98 x 1 | 120 / 33.2 $^a$ | 68 |
| 27 | BJH | 0.87 x 0.87 x 3 | 100 / 13.6$^a$ | 0.96 x 0.96 x 3 | 140 / 10.0 $^a$ | 47 |
| 28 | BJH | 0.68 x 0.68 x 2 | 100 / 9.0$^a$ | 0.66 x 0.66 x 2 | 100 / 7.0 $^a$ | 61 |
| 29 | BJH | 0.75 x 0.75 x 2 | 120 / 16.6$^a$ | 0.73 x 0.73 x 2 | 100 / 18.3 $^a$ | 58 |
| 30 | BJH | 0.66 x 0.66 x 2 | 100 / 13.5$^a$ | 0.66 x 0.66 x 2 | 100 / 12.2 $^a$ | 84 |

## 3.2. Use

The usage instructions and sample code are available at https://github.com/deshanyang/Liver-DIR-QA. For visualizing the landmark pairs within the dataset, we provide our MatchGui MATLAB tool. This program enables researchers to view, manipulate, and annotate landmark pairs as needed.

The images in the dataset originated from various scanners and had generally good image quality, high resolution, and enhanced vessel contrast. To simulate more realistic clinical scenarios, researchers can down-sample the images and add noise before testing their deformable image registration methods. This will allow more comprehensive evaluations of DIR performance under more challenging conditions.

## 4. Discussion

In this work, we have developed a comprehensive landmark pair dataset for the liver using patients' CT images. The 2028 landmarks in 30 image pairs, with an average of 68 landmark pairs per case, represent a substantial improvement compared to previous reports, which typically only included about 5-15 landmark pairs per case.[35,37] Furthermore, since the process of detecting these landmarks was semi-automated, it minimized inter-observer variations and ensured consistent landmark placements. A thorough manual verification process was applied to ensure the accuracy and correspondence of the final landmark pairs. As demonstrated in Figure 2, the distribution of these landmarks was notably uniform across the liver, which is desirable for image registration verification. We estimated the positional accuracy of our landmarking procedure to be approximately 0.64 ± 0.40 mm.

It is worth noticing that assessing the precision of landmark pairs posed challenges due to the absence of ground truth. Previous manual landmarking studies[21] evaluated landmark positional uncertainty by measuring inter-observer variations. One advantage of our semi-automatic procedure was to avoid the manual annotation process and the associated uncertainty. This was demonstrated by the fact that the average TRE (0.64 mm) calculated using the digital phantom simulation were considerably lower compared to the inter-observer uncertainty (0.89 mm) reported in previous DIRLAB 4DCT dataset, which were established on organs with even stronger soft tissue contrast.[21]

As our landmark pairs were initially placed using an image registration algorithm pTVreg,[48] there might be a bias toward it or its similar algorithms. Future efforts should integrate different or multiple algorithms into the landmarking workflow to mitigate this potential bias. Based on this thought, we plan to conduct an inter-algorithm uncertainty study, which may offer a more reasonable assessment of the quality of this landmark pair library, as opposed to relying on inter-observer uncertainty studies for evaluating landmark correspondence.

In the future, we plan to further enhance the vessel segmentation step using deep learning networks. With a more robust liver vessel segmentation method, we could add more landmark pairs to our benchmark dataset and reduce the fault pairs. The approach established in this research could be applied to other anatomical regions where vascular structures could be reasonably segmented. Ongoing efforts in our group are focused on creating similar datasets for the lung, abdomen, and head-neck regions. Additionally, we also plan to develop a deep learning classifier to reject landmark outliers introduced during the landmark detection process. This enhancement could further automate the workflow, thereby eliminating the manual verification step, which is currently the most time-intensive step of the process. Such progress could pave the way for the generation of even larger datasets and enable the application of our workflow for patient-specific DIR quality assurance.

## 5. Conclusions

In this study, we created a rich library of landmark pairs for the liver on 30 pairs of clinical CT scans. This dataset library can serve as a valuable resource for validating and advancing DIR algorithms on liver CTs. The data is publicly accessible, allowing researchers to independently assess the performance of their own DIR algorithms.

## 6. Acknowledgments

This research was supported by the National Institute of Biomedical Imaging and Bioengineering (NIBIB) grant R01-EB029431. The results here are in whole or part based upon data generated by the TCGA Research Network.